\documentstyle[aps,epsf,multicol,prb]{revtex}

\draft

\begin{document}

\title{Shear-induced vortex decoupling in Bi$_2$Sr$_2$CaCu$_2$O$_8$
crystals}
\author{B. Khaykovich, D. T. Fuchs\cite{byline}, 
K. Teitelbaum, Y. Myasoedov, and E. Zeldov}
\address{Department of Condensed Matter Physics,  
Weizmann Institute of Science, 76100 Rehovot, Israel}
\author{T. Tamegai and S. Ooi\cite{ooi}}
\address{Department of Applied Physics, The University of Tokyo, 
Hongo, Bunkyo-ku, Tokyo 113-8656, Japan}
\author{M. Konczykowski} 
\address{CNRS, UMR 7642, Laboratoire des Solides Irradies, 
Ecole Polytechnique, 91128 Palaiseau, France}
\author{R. A. Doyle and S. F. W. R. Rycroft}
\address{IRC in Superconductivity, University of Cambridge, 
Cambridge CB3 0HE, United Kingdom}
\date{\today}
\maketitle

\begin{abstract}
Simultaneous transport and magnetization studies in 
Bi$_2$Sr$_2$CaCu$_2$O$_8$ crystals at elevated 
currents reveal large discrepancies, including
{\em finite} resistivity at temperatures of 40K {\em below} 
the magnetic irreversibility line. This resistivity,
measured at the top surface, 
is non-monotonic in temperature and extremely non-linear. 
The vortex velocity derived from magnetization is six 
orders of magnitude lower than the velocity derived from 
{\em simultaneous} transport measurements.
The new findings are ascribed to a shear-induced 
decoupling, in which the pancake vortices flow only in the top 
few CuO$_2$ planes, and are decoupled from the pinned vortices in 
the rest of the crystal. 
\end{abstract}

\begin{multicols}{2}
Transport measurements are one of the most common methods to study
vortex dynamics in superconductors. The derived resistivity
$\rho$ describes the vortex motion as a function of temperature,
field, and the applied current. In high-temperature
superconductors the situation is more complicated due to their high
anisotropy and layered structure. The corresponding resistivity has two
main components, the in-plane resistivity $\rho_{ab}$ and the
out-of-plane $\rho_c$, with typical ratio of $\rho_c / \rho_{ab} \simeq
10^{4}$ in the normal state of Bi$_2$Sr$_2$CaCu$_2$O$_8$ (BSCCO)
crystals \cite{Busch}. As a result, the measured resistance
$R$ is a non-trivial function of sample geometry and contact
configuration \cite{Busch,safar_prb}, which is further significantly
complicated by the nonlinear current dependence of $\rho_{ab}$ and
$\rho_c$. Yet it is generally assumed that the 
physical mechanism that governs the dissipation can be described in terms
of current density, namely, the local values of $\rho_{ab}$ and
$\rho_c$ are determined \cite{Brandt_rev,Koshelev} 
by the corresponding in-plane and out-of-plane
current densities $j_{ab}$ and $j_c$ (ignoring possible nonlocal
effects \cite{Rev,Huse}). In this paper we demonstrate that in highly
anisotropic materials like BSCCO this assumption may not be valid. We find
that at elevated currents an additional new term, the c-axis {\em gradient} of
the in-plane current $dj_{ab}/dz$, becomes the dominant parameter in the description of the
local dissipation. This current gradient
induces large velocity gradients $dv_{ab}/dz$ of the pancake vortices in
the different CuO$_2$ planes, leading to their decoupling and to corresponding
dramatic increase in $\rho_c$ and anisotropy
$\gamma$. In the presence of inhomogeneous currents, this mechanism
results in fundamental changes in the transport behavior, including
large measurable reentrant resistance well {\em below} the magnetically
determined irreversibility line (IL).

The studies were carried out on several high quality BSCCO
crystals \cite{crystal_grow}, with $T_c \simeq 90$ K. Four wires
were attached to the gold pads evaporated on freshly cleaved top {\em ab} surfaces, as shown
schematically in the inset to Fig. 1a. The bottom surface of the
crystals, free of electrical contacts, was attached to an array of 19
2DEG Hall sensors \cite{Hall_sensors}, $30\times30$ $\mu m^2$ each,
allowing {\em simultaneous} resistance $R$ and local magnetization
measurements in the presence of transport current. In addition, several
crystals were irradiated by very low doses of 5.8 GeV Pb ions to produce
columnar defects with concentrations corresponding to matching fields of
$B_{\phi} $ = 5, 20, and 60 G. In order to focus on bulk vortex dynamics
and to avoid complications due to surface barriers \cite{dandan_nature},
most of the samples were prepared in a form of large square platelets
with electrical contacts positioned far from the edges \cite{cut}. Some
of the crystals were cut into strip shape for comparison. Although
detailed behavior varies from sample to sample, the qualitative features
described below were observed in all the samples
regardless of the surface barriers and the irradiation doses. Here we
present data for as-grown crystal A of size
$1700\times1300\times10 \mu m^3$, which was subsequently cut into a
strip, and for an irradiated crystal B $1000\times700\times30 \mu m^3$
with $B_{\phi}$ = 60 G.

Our main general observation is displayed in Fig. 1. Figure 1a 
shows the $R(T)$ of the unirradiated crystal A 
at various applied fields $H_a\|$ c-axis at elevated applied current 
of 30 mA. At lower currents, that are usually used in 
transport studies, $R(T)$ behaves monotonically with temperature 
as shown for comparison by the dashed line in Fig. 1a for 10 mA 
at 500 Oe. However, when $I_a$ is increased we find surprisingly a
very pronounced reentrant behavior, in which $R(T)$ reaches a minimum
at some characteristic temperature $T_{min}$,
but then increases again, often by more than an order of magnitude, 
at lower $T$. At 100 Oe a sharp drop in $R(T)$ is observed at the first-order
melting transition (FOT), $T_m$, followed
\begin{figure}
\begin{center}
\begin{minipage}{3.4in}
\setlength{\epsfxsize}{3in}
\leavevmode
\epsfbox{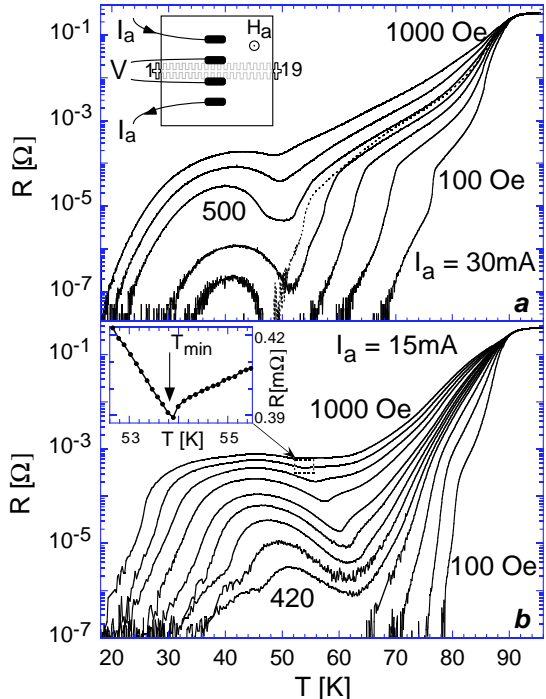}
\caption{Resistance (log scale) vs. temperature at various applied
fields. (a) As-grown sample A at 100, 200, 300, 400, 500, 700, and
1000 Oe; $I_a=30$ mA. (b) Irradiated sample B at 100, 200, 300, 380,
420, 460, 500, 550, 600, 700, 800, 900, and 1000 Oe; $I_a=15$ mA.
Inset to (b) shows $R(T)$ in the vicinity of $T_{min}$ at 900 Oe on
linear scale. }
\label{fig1}
\end{minipage}
\end{center}
\end{figure}
by a monotonic resistive tail in the quasi-ordered-lattice phase, as described 
previously \cite{cut}. At 200 Oe the behavior is still monotonic 
similar to 100 Oe data. At 300 Oe, however, $R(T)$ decreases monotonically 
down to our noise level at about 55 K, but then reappears again between 
46 and 34 K. This reentrant behavior is strongly pronounced at 400 Oe 
and up to fields above 1000 Oe. In addition, a sharp discontinuity in the
derivative $dR(T)/dT$ is often observed at $T_{min}$ as shown in the inset to Fig. 1b.
The irradiated samples display essentially similar properties, as shown 
in Fig. 1b, with a few consistent 
differences: (i) The resistive kink at the FOT is smeared by the low 
dose of columnar defects, consistent with previous studies 
\cite{disorder}, (ii) $T_{min}$ is generally shifted to higher 
temperatures, and (iii) the anomalous behavior appears at 
lower currents, as shown by the 15 mA data in Fig. 1b. 
The general form of $R(T)$ in Fig. 1 is somewhat reminiscent of the
``peak effect'' observed \cite{NbSe2} in NbSe$_2$ near $H_{c2}$, or at
the melting transition \cite{YBCO-peak-eff}
in YBa$_2$Cu$_3$O$_7$.
The behavior reported here, however, is much broader and is of a fundamentally
different nature.

In the unirradiated crystals 
the reentrant features usually appear above 25 mA, close to our upper 
bound of 30 mA, limited by the contact resistance. 
The enhanced pinning in the irradiated samples causes the anomalous
behavior to begin at $I_a$ as low as 10 mA, allowing a more systematic study.
Figures 2a and 2b thus present $R(T)$
\begin{figure}
\begin{center}
\begin{minipage}{3.4in}
\setlength{\epsfxsize}{3in}
\leavevmode
\epsfbox{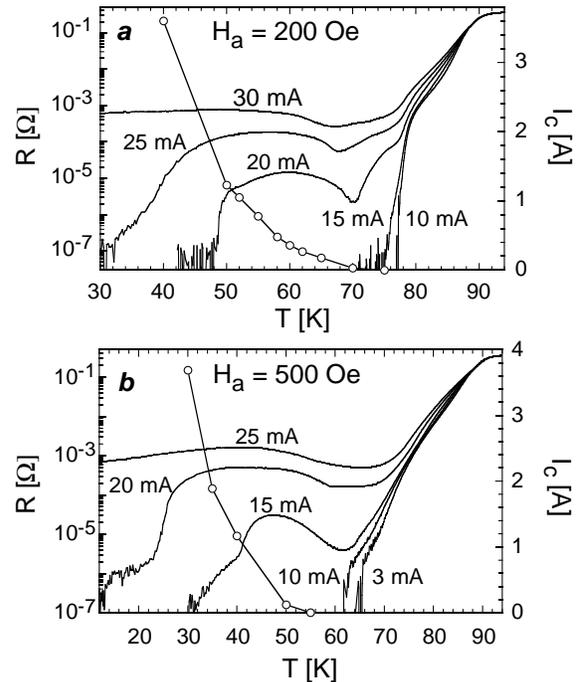}
\caption{Resistance at various $I_a$ (left axis, log scale) and 
magnetically measured critical current (right axis, linear scale, 
open circles) vs. $T$ for the irradiated sample, $H_{a}=$ 200 Oe (a) 
and $H_{a}=$ 500 Oe (b).}
\label{fig2}
\end{minipage}
\end{center}
\end{figure}of crystal B at various $I_a$ at 200 and 500 Oe respective-
ly. At low currents $R(T)$ is monotonic and only weakly current dependent,
and has a rather well-defined temperature $T_{R=0}$ at which 
$R(T)$ drops below experimental resolution. All these qualities change 
drastically at higher currents. The appearance of non-monotonic $R(T)$ 
is accompanied by extreme non-linearity. An increase of $I_a$ by 30\% 
or less may result in enhancement of $R$ by orders of magnitude. 
Furthermore, the same current increase causes a decrease of $T_{R=0}$ 
by tens of degrees K.

In order to elucidate the origin of this anomalous behavior we have 
carried out simultaneous local magnetization measurements
{\em in presence} of $I_a$ as shown in Fig. 3a for 
$T=30$ K. The results seem to be paradoxical. 
Finite resistivity should be present only above the magnetically 
measured IL, which in Fig. 3a occurs above 1600 Oe. 
Below the IL the resistivity should be immeasurable in 
standard transport measurements. This is indeed the case at low 
$I_a$. However, at elevated currents, substantial $R$
is measured {\em concurrently} with the hysteretic magnetization well
below the IL (in Fig. 3a). Figure 3b shows the corresponding 
field profile $B_z(x)$ obtained by the Hall sensors at 400 Oe on
increasing and decreasing $H_a$ in presence of $I_a$. A clear Bean profile
is observed \cite{Eli_europhys}. 
Fitting this profile to the theoretical $B_z(x)$ in platelet 
sample \cite{zeldov-clem} results in critical current density 
$J_c = 2\times10^4$ $A/cm^2$, which translates into total critical 
current of $I_c = 4.2$ A. Obviously, a transport current of 25 mA, 
which is less than 1\% of $I_c$, cannot result in measurable resistance
in the standard critical state models.
\begin{figure}
\begin{center}
\begin{minipage}{3.4in}
\setlength{\epsfxsize}{3in}
\leavevmode
\epsfbox{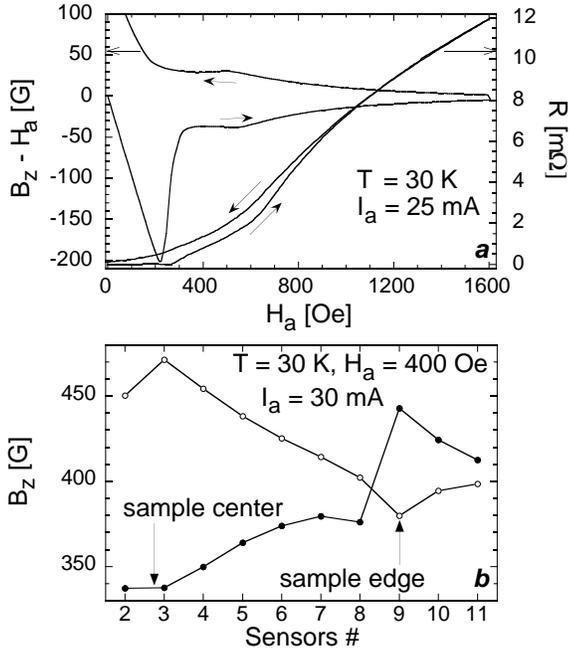}
\caption{ (a) Resistance (right axis) and hysteretic magnetization loop 
in the sample center (left axis) vs. $H_a$ at $T=30$ K and 
$I_a=25$ mA for crystal B. (b) Profile of magnetic induction across 
the sample at 400 Oe on increasing ($\bullet$) and decreasing 
($\bigcirc$) fields.}
\label{fig3}
\end{minipage}
\end{center}
\end{figure}
Figure 2 shows $I_c(T)$ determined from the Bean pro-
files together with the $R(T)$ data. The reentrant resistance always occurs in 
the region where $I_a\ll I_c$. We can also compare the electrical 
fields. In the magnetization measurements of Fig. 3 the slow sweep
of $H_a$ of about 0.3 Oe/sec induces an average electrical field of $4\times10^{-10}$ V/cm in the sample, which 
corresponds to a very low vortex velocity of 10$^{-4}$ cm/sec 
at 400 Oe, for example. The resistively  measured electrical field, 
induced by the simultaneous transport current, on the other 
hand, is $1.4\times10^{-3}$ V/cm. This translates into vortex 
velocity of 350 cm/sec, which is 6 orders of magnitude higher. 
Clearly existence of such a high vortex flow rate induced applied 
by {\em concurrently} $I_a$ should have resulted in completely 
reversible magnetization, in sharp contrast to the data.
So paradoxically we have a situation in which a transport current, 
which is two orders of magnitude {\em lower} than $I_c$, results in 
vortex velocity which is 6 orders of magnitude {\em higher} than the 
flux creep vortex velocity at $I_c$.

We emphasize that we do not observe any heating effects as confirmed by
monitoring the location of the resistive transitions at $T_c$ and $T_m$
at various currents. Furthermore, and more importantly, since the two
measurements are carried out simultaneously, the transport and the
magnetization results should be affected by heating to the same
extent, and therefore heating effects can be ruled out as a possible
explanation of the observed large discrepancy between the two
measurements.

We describe the observed phenomena in terms of shear-induced decoupling
and shear-enhanced anisotropy. At elevated temperatures in the 
vortex-liquid phase the in-plane current density $j_{ab}(z)$ decreases 
approximately exponentially from the top surface to the bottom 
\cite{Busch}. In samples of typical dimensions the current at the bottom 
is about four orders of magnitude lower than at the top,
due to high anisotropy $\gamma$ of BSCCO. As $T$ is 
decreased the pinning becomes more effective. At some temperature 
the current density at the bottom becomes comparable or lower than 
$J_c$. As a result the vortices at the bottom stop moving or reduce 
their velocity substantially, whereas the vortices at the top maintain
their high velocity since the current density there is significantly 
above $J_c$. As a result, the velocity gradient between the planes 
$dv_{ab}/dz$ is increased. The thickness of the vortex pinned layer 
grows as $T$ is decreased resulting in progressively larger velocity 
gradients within the vortex mobile part at the top. The enhanced 
$dv_{ab}/dz$ results, in turn, in shear-induced phase slippage between
the adjacent CuO$_2$ planes reducing the Josephson coupling and leading
to decoupling of the planes \cite{Brandt_rev,Koshelev}. As a result 
the $\rho_c$, and hence the measured $R$, are increased significantly
causing the reentrant $R(T)$. This process is highly nonlinear, 
since larger $I_a$ results in larger 
$dj_{ab}/dz$, causing an increase in $\rho_c$ and $\gamma$, 
which lead in turn to even shallower current profile and thus extremely
nonlinear $R(I)$. Since the 
onset of the enhanced shear is induced by pinning, low dose irradiation
causes the anomalous behavior to set in at higher $T$ and lower $I_a$ as seen in Fig. 1. 
It is important to realize that the enhanced $\rho_c$ prevents the lower
vortex pinned region from effectively shunting the current. In the case of
full decoupling, for example, $\rho_c$ becomes comparable to the normal
state $\rho_c$, which in BSCCO crystals of typical dimensions results
in about $10^{-4}$ $\Omega$ resistance between two adjacent CuO$_2$ planes.
Thus the serial c-axis resistance of several decoupled layers in the upper
part of the sample can be of the order of $10^{-3}$ $\Omega$, thus
effectively insulating the lower zero-resistance region. Any velocity
gradient $dv_{ab}/dz$ should, in principle, completely decouple the layers,
since the time-averaged Josephson phase difference reduces to zero.
More detailed analysis \cite{clem-decoupling}, however, shows that the
pancake shear occurs through a `stick - slip' process, which retains a
finite Josephson coupling at low $dv_{ab}/dz$. Thus the Josephson current
between the layers can be sustained at low $dv_{ab}/dz$, and is lost at
higher rates of pancake shear.

The described mechanism is {\em conceptually} different from conventional 
electrodynamics in which the dissipation is determined only by the 
local vortex velocity, and not by the velocity {\em gradient}.
The underlying physics is also very different from the process of nonlocal
resistivity \cite{Rev,Huse}, in which a force applied to a pancake vortex
results in a nonlocal drag of pancakes in adjacent layers. This drag, which
is proportional \cite{Huse} to $d^2v_{ab}/dz^2$, induces a more uniform vortex
flow, in contrast to the described shear-induced decoupling which has an opposite
effect of amplification of the velocity differences.

Since in most of the crystal thickness the vortices are
\begin{figure}
\begin{center}
\begin{minipage}{3.4in}
\setlength{\epsfxsize}{2.5in}
\leavevmode
\epsfbox{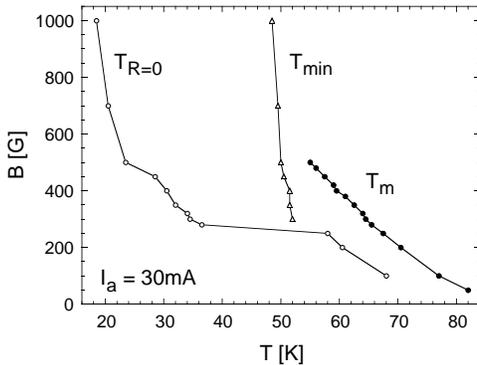}
\caption{$B-T$ phase diagram of crystal A at 30 mA. $T_m$ ($\bullet$) is 
the first-order melting line. $T_{min}$ ($\bigtriangleup$) is 
the position of the resistivity minimum, and $T_{R=0}$ ($\bigcirc$) 
is the temperature at which $R$ falls below the experimental resolution.}
\label{fig4}
\end{minipage}
\end{center}
\end{figure}pinned, Bean profiles are obtained in the bulk upon sweeping $H_a$ (Fig. 3b). At the same 
time the transport current confined to the top planes is sufficient 
to create vortex flow in these planes, resulting in a measurable 
voltage drop, and explaining the apparent discrepancy between the
transport and magnetization data. A similar mechanism probably occurs when the 
c-axis properties are probed by transverse ac susceptibility and 
transport in the zero-field-cooled state as reported previously 
\cite{rodriguez}. Our model also explains the hysteretic
$R(H)$ in Fig. 3a. The field $B$ in the top planes is determined 
by the hysteretic Bean profile in the rest of the crystal. As a result, 
the measured $R(H_a)$ displays hysteresis, even though the vortices are mobile in the top planes. Consequently, 
this apparent hysteresis practically disappears when $R$ 
is plotted vs. the average field $B$ measured by the sensors, rather 
than vs. $H_a$.

Figure 4 summarizes the $B-T$ diagram of sample A at 30 mA. A similar
diagram is obtained for the irradiated 
crystal B. The resistance vanishes within our resolution at $T_{R=0}$, 
the position of which is highly current-dependent. The resistively 
determined FOT extends 
to fields of about 500 Oe in this sample.
The anomalous reentrant $R(T)$ is observed
in the broad region between $T_{R=0}$ and $T_{min}$, where 
vortices are pinned in the bulk. It is interesting to note that the decoupling 
process occurs first in the entangled vortex-solid phase above 500 Oe, 
where vortex cutting is apparently relatively easy. As $I_a$ is increased, the
shear-induced decoupling expands also into the 
quasi-ordered-lattice phase below $T_m$ as is the case in Fig. 4.
The non-linearities expand also partially into the liquid phase above
$T_{min}$ and $T_m$, where the shear process can still modify $\rho_c$
at higher currents.

One can evaluate the number of the CuO$_2$ planes participating in the
flow of vortex pancakes. In Fig. 2b at 30 K, for example,
the total critical current determined from the Bean profiles is
$I_c~=~3.7~A$. This translates into critical current of about 0.18 mA
per (double) CuO$_2$ plane of the sample. Thus, the top 140 planes would
be sufficient to carry the entire $I_a$=25mA without observable dissipation.
In order to obtain the large measured resistance the actual current in the
planes has to be significantly above 0.18 mA.
Furthermore, the lower part of the sample also carries a substantial
portion of $I_a$. We therefore conclude that only a few tens of CuO$_2$
planes at the top of the crystal participate in the described process.
Current-induced decoupling was recently analyzed theoretically
in absence of Josephson coupling \cite{clem-decoupling}. It was found
that above some critical value of the current applied to the top plane,
slippage between magnetically coupled planes occurs, leading to an abrupt
increase in the vortex velocity in the top plane. Since in this
model $\rho_c$ is infinite, there is no corresponding current redistribution.
Computer simulations \cite{Stroud-Dominguez} of Josephson-coupled planes also show sharp
current-induced decoupling. Vortex cutting was also observed in YBCO
crystals at elevated currents in flux transformer measurements
\cite{safar_prl,Lopez}. The much lower anisotropy of YBCO, however,
causes only a small change in the transport properties. In BSCCO in contrast
the shear-induced decoupling
leads to variations by orders of magnitude in the observed behavior.

In conclusion, we reveal large apparent discrepancies between simultaneously measured 
magnetization and transport properties, which reveal an important 
feature of vortex dynamics in the presence of inhomogeneous currents. In layered 
superconductors at elevated currents the c-axis resistivity becomes a
strong function of the local current gradient $dj_{ab}/dz$, instead of being determined just 
by the current density. This mechanism results in a highly nonlinear
reentrant resistance well below the irreversibility line.

This work was supported by Israel Ministry of Science, by US-Israel 
Binational Science Foundation (BSF), by CREST, and by the Grant-in-Aid 
for Scientific Research from the Ministry of Education, Science, Sports, 
and Culture, Japan. 

\end{multicols}
\end{document}